\documentclass[amsmath,amssymb,superscriptaddress,nobalancelastpage,aps,twocolumn,showpacs]{revtex4-1}

\usepackage{array,multirow,graphicx}
\usepackage{braket}
\usepackage{epstopdf}
\usepackage{graphicx}
\usepackage{siunitx}
\usepackage{nicefrac}
\usepackage[normalem]{ulem}
\usepackage{varioref}
\usepackage{xcolor}
\usepackage{xfrac}
\usepackage{xr-hyper}
\usepackage{hyperref}

\definecolor{bl}{rgb}{0.0,0.2,0.6}

\hypersetup{colorlinks,linkcolor=blue,urlcolor=blue,citecolor=blue}

\newcommand{\PLCCO}{Pr$_{1.3-x}$La$_{0.7}$Ce$_x$CuO$_4$}
\newcommand{\NCCO}{Nd$_{2-x}$Ce$_x$CuO$_4$}

\begin{document}
	\author{M.~Miyamoto}
	\affiliation{Institute for Solid State Physics, The University of Tokyo, Kashiwa, Chiba 277-8581, Japan}
	
	\author{M.~Horio}
	\email{mhorio@issp.u-tokyo.ac.jp}
	\affiliation{Institute for Solid State Physics, The University of Tokyo, Kashiwa, Chiba 277-8581, Japan}
	
	\author{K.~Moriya}
	\affiliation{Department of Engineering and Applied Sciences, Sophia University, 7-1 Kioi-cho, Chiyoda-ku, Tokyo 102-8554, Japan}
	
	\author{A.~Takahashi}
	\affiliation{Department of Applied Physics, Tohoku University,	6-6-05 Aoba, Aramaki, Aoba-ku, Sendai 980-8579, Japan}
	
	\author{K.~Tanaka}
	\affiliation{UVSOR Synchrotron Facility, Institute for Molecular Science, Okazaki 444-8585, Japan}
	
	\author{Y.~Koike}
	\affiliation{Department of Applied Physics, Tohoku University,	6-6-05 Aoba, Aramaki, Aoba-ku, Sendai 980-8579, Japan}
	
	\author{T.~Adachi}
	\affiliation{Department of Engineering and Applied Sciences, Sophia University, 7-1 Kioi-cho, Chiyoda-ku, Tokyo 102-8554, Japan}
	
	\author{I.~Matsuda}
	\affiliation{Institute for Solid State Physics, The University of Tokyo, Kashiwa, Chiba 277-8581, Japan}
	
	\title{Differentiation of electron doping and oxygen reduction in electron-doped cuprates}
	
\begin{abstract}
Electron-doped cuprates require not only electron doping by chemical substitution 
but also post-growth reduction annealing for realizing superconductivity. 
However, electron concentration can also be varied by reduction annealing, making it challenging 
to disentangle the respective influences of electron concentration and oxygen non-stoichiometry. 
Here, by combining alkali-metal dosing and angle-resolved photoemission spectroscopy, we 
monitored changes in the electronic structure of an electron-doped cuprate while supplying 
additional electrons to its surface without modifying oxygen content. Whereas a Fermi surface 
reconstruction due to long-range antiferromagnetic order was suppressed by alkali-metal 
deposition, the pseudogap -- which is associated with short-range spin/charge correlations and 
can be suppressed by efficient reduction annealing -- was found to persist. The results highlight 
significant contribution of impurity oxygen atoms to pseudogap formation in electron-doped 
cuprates.
\end{abstract}
	
\maketitle

\section{Introduction}

Identifying the condition for the emergence of superconductivity offers essential understanding 
on the origin of unconventional superconductivity. In electron-doped cuprate superconductors 
$R_{2-x}$Ce$_x$CuO$_4$ ($R$: rare earth), electron doping to the parent antiferromagnetic (AF) Mott insulator 
via Ce substitution alone is insufficient to induce superconductivity. Additional post-growth 
annealing in a reducing atmosphere is required~\cite{TokuraNature1989,TakagiPRL1989}. 
Neutron-scattering studies found that impurity oxygen atoms reside at the apical site [Fig. ~\ref{character}(a)] in as-grown 
samples and are partially removed by reduction annealing~\cite{RadaelliPRB1994,SchultzPRB1996,FujitaJPSJ2021}. 
It has been demonstrated that improving the efficiency of oxygen reduction extends the superconducting 
dome in the phase diagram toward lower Ce concentration~\cite{BrinkmannPRL1995,AdachiJPSJ2013}. 
In fact, in thin films, where the large surface-to-volume ratio facilitates oxygen diffusion, 
superconductivity has been realized even without any Ce substitution~\cite{TsukadaSSC2005,MatsumotoPhysicaC2009}. 
On the other hand, angle-resolved photoemission spectroscopy (ARPES) studies on efficiently annealed 
samples found significantly larger electron concentration than the nominal Ce concentration
~\cite{HorioNatCommun2016,SongPRL2017,HorioPRB2018,LinPRR2021,MiyamotoPRB2025,SongNatCommun2025}. 
This observation suggests that reduction annealing removes oxygen atoms not only from the impurity 
apical site but also from the regular sites such as the CuO$_2$ planes and the $R_2$O$_2$ layers [Fig.~\ref{character}(a)]. 
Consequently, both the impurity and doped electron concentrations could vary concomitantly upon 
reduction annealing, making it difficult to disentangle the relevance of those two factors with 
superconductivity.\\
\indent The most straightforward way to examine the influence of electron doping would be changing 
Ce concentration. However, according to thermogravimetric analysis on \NCCO, the amount of oxygen 
evacuation by reduction annealing ($\delta$) considerably decreases with the increase of Ce contents; 
from $\delta = 0.04$--0.07 at $x=0$ to $\delta \sim 0.01$ at $x=0.15$~\cite{TakayamaPhysicaC1989,SuzukiPhysicaC1990,KimPhysicaC1993}, 
implying that the apical-site occupation depends also on Ce concentration. 
In order to fully separate the two degrees of freedom -- the concentrations of doped 
electrons and impurities -- an alternative approach is thus demanded. Recently, alkali-metal 
deposition has been increasingly recognized as an effective method to dope electrons to the 
surface of strongly correlated materials~\cite{HossainNature2008,KimScience2014,MiyataNatMater2015,RenSciAdv2017,HuNatCommun2021,HorioComPhys2023}. 
Since alkali-metal atoms merely adsorb on the surface, inherent oxygen impurity level of the sample 
should be unaffected. By combining that treatment with surface-sensitive ARPES measurements, one 
could selectively target the effect of electron doping on the electronic structure.\\
\indent In the present work, we dosed potassium (K) on the as-grown samples of the electron-doped cuprate 
\PLCCO\ (PLCCO) and investigated changes in the electronic structure by ARPES. Even after 
extensive electron doping to the surface, a pseudogap arising from short-range spin/charge 
correlations was found to persist. This was in contrast to the virtually full suppression achieved by 
efficient reduction annealing. Considering the known competition between the pseudogap and 
superconductivity, the results highlight the critical role of impurity oxygen removal in shaping 
superconductivity of electron-doped cuprates.

\section{Methods}
Single crystals of PLCCO ($x = 0.08, 0.10$) were synthesized by the traveling-solvent floating-zone 
method~\cite{AdachiJPSJ2013}. The as-grown samples did not show superconductivity. Post-growth 
reduction annealing was conducted for the $x = 0.10$ sample. The annealing process consists of 
three steps; protect annealing~\cite{AdachiJPSJ2013}, low temperature annealing~\cite{KrockenbergerSciRep2013}, 
and dynamic annealing~\cite{WangPRB2009,SumuraJPSCS2018} at temperatures of 800~$^\circ$C, 400~$^\circ$C, 
and 500~$^\circ$C, respectively. After the sequential annealing, the $x = 0.10$ sample showed 
superconductivity with $T_\mathrm{c} = 27.8$~K.\\
\indent ARPES measurements were performed at BL5U of UVSOR with the total energy resolution 
better than 30~meV. Circularly polarized, $h\nu$ = 55~eV incident light was used. The samples 
were cleaved and measured \textit{in situ} at $T=20$~K under the pressure better than $6 \times 
10^{-11}$~mbar. K deposition was conducted in the ultra high vacuum chamber by commercial 
dispensers (SAES Getters S.p.A.) while the sample temperature was kept at 20 K. The energy distributions curves 
(EDCs) shown hereafter have been normalized to the intensity either at $0.4$ eV [Fig. \ref{character}(c) and Figs. \ref{Kdosed}(e)-(f)] 
or $0.1$ eV [Fig. \ref{annealed}(d)] below the Fermi level, after subtracting a momentum-independent background~\cite{KaminskiPRB2004,MattPRB2015}. 
The background was constructed by averaging and smoothing EDCs taken near ($\pi$, $\pi$), far from the Fermi wavenumber.

\section{Results}
\begin{figure}[ht]
	\begin{center}
		\includegraphics[width=0.48\textwidth]{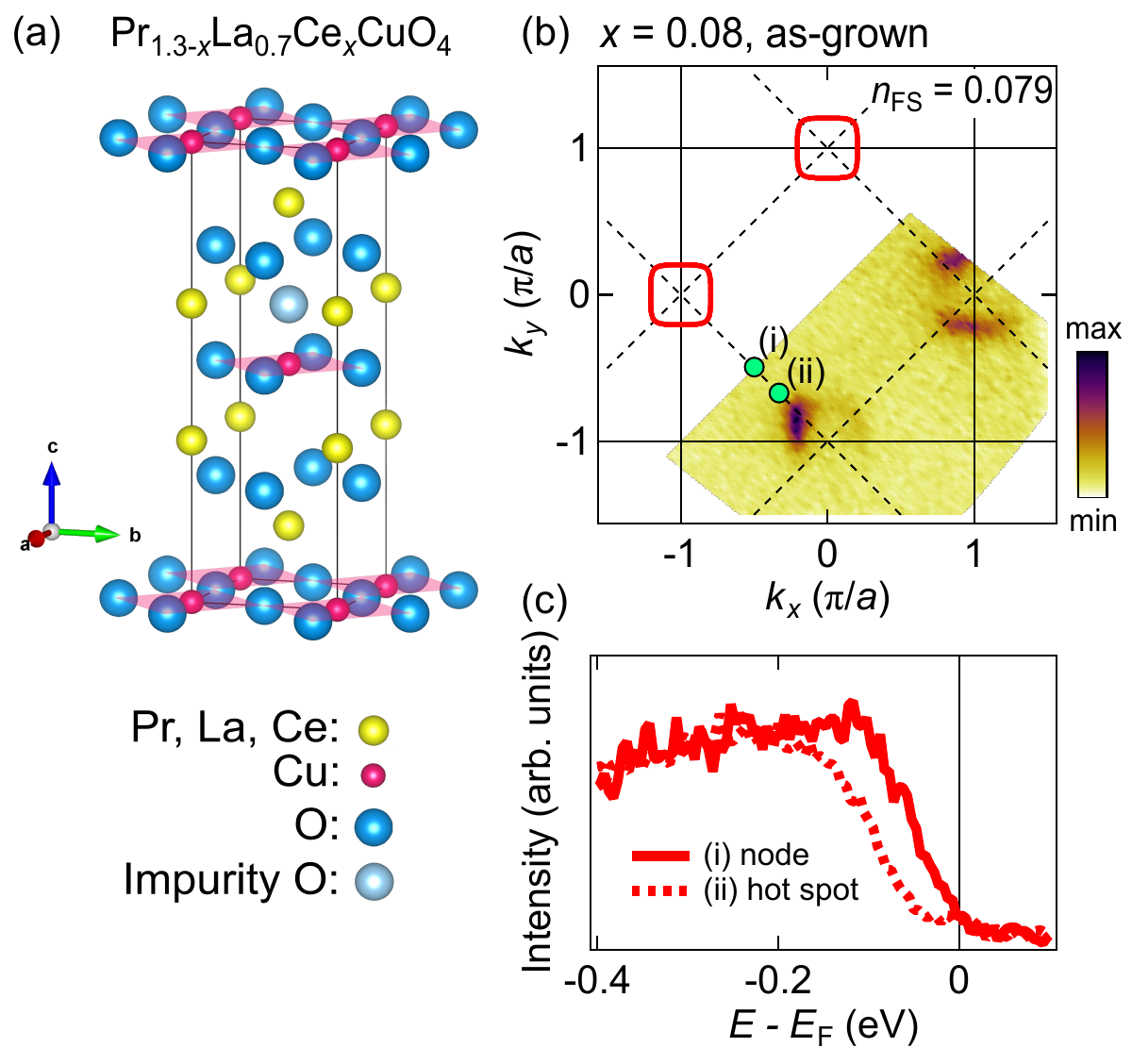}
	\end{center}
	\caption{\textbf{Characterization of the as-grown sample.} (a) Crystal structure of PLCCO with an impurity 
	oxygen atom at the apical site. (b) Fermi surface map of the as-grown PLCCO ($x=0.08$) sample. The red curves indicate the result 
	of tight-binding fitting with the $\sqrt{2} \times \sqrt{2}$ AF order (eq.~\ref{TB_AFPG}). 
	(c) EDCs at the node and the hot spot for the as-grown PLCCO ($x=0.08$) sample taken from the points 
	shown in (b). 
	} 
	\label{character}
\end{figure}

Prior to investigating the effect of K adsorption on the electronic states near the Fermi level, 
the pristine surface of the as-grown PLCCO ($x=0.08$) sample was first characterized, whose Fermi 
surface map is shown in Fig.~\ref{character}(b). As expected for as-grown electron-underdoped 
cuprates in the long-range AF ordered state, hole-like Fermi surfaces centered at the Brillouin-zone 
(BZ) corners are reconstructed into electron-like Fermi surfaces around ($\pi$, 0) and equivalent 
momentum points~\cite{ArmitagePRL2002}. A large gap is observed in the EDCs at the node 
around ($\pi /2$, $\pi /2$) and the hot spot, where the original unreconstructed 
Fermi surface intersects the $\sqrt{2} \times \sqrt{2}$ AF BZ boundary [Fig.~\ref{character}(c)]. 
The reconstructed Fermi surface can be modeled by the following the tight-binding model
~\cite{ParkPRB2007,IkedaPRB2009} of the square lattice with the $\sqrt{2} \times \sqrt{2}$ AF order,
\begin{align}
	\epsilon - \mu &= \epsilon _0 \pm \sqrt{\Delta^2 + 4t^2 (\mathrm{cos} \ k_xa + \mathrm{cos} \ k_ya)^2}  \nonumber \\ 
	& -4t'\mathrm{cos} \ k_xa \ \mathrm{cos} \ k_ya -2t''(\mathrm{cos} \ 2k_xa + \mathrm{cos} \ 2k_ya), \label{TB_AFPG} 
\end{align}
where $t$, $t'$, and $t''$ are transfer integrals between the nearest-neighbor, second-nearest-neighbor, 
and third-nearest-neighbor Cu sites, respectively. $\epsilon _0$ denotes the energy of the band 
center relative to the chemical potential $\mu$. 2$\Delta$ represents the potential difference 
between the two spin sublattices that opens a spectral gap. In the course of Fermi surface fitting, 
$\Delta$ was fixed at 0.15~eV referring to a previous study \cite{HorioPNAS2025}. 
Also, as widely assumed for cuprates~\cite{YoshidaPRB2006,IkedaPRB2009,HorioNatCommun2016}, 
$t''/t'$ was fixed at -0.5. Fermi surface fitting conducted under those constraints successfully 
reproduces the observed Fermi surface as shown in Fig.~\ref{character}(b) with the parameters $t/t' = -0.17$ 
and $\epsilon_0/t = 0.05$. From the area of the modeled Fermi surface, the electron concentration 
was estimated to be $n_\mathrm{FS} = 0.079 \pm 0.010$ per Cu atom, which is consistent with the Ce concentration.\\
\begin{figure*}[ht]
	\begin{center}
		\includegraphics[width=0.98\textwidth]{fig/Fig2_v8.pdf}
	\end{center}
	\caption{\textbf{Effect of electron doping on the as-grown sample.} (a) Schematic illustration of photoemission 
	from the K-dosed surface. The adsorbed K atoms are ionized and donate electrons 
	to the sample surface. 
	(b) Core-level spectra of the as-grown PLCCO ($x=0.08$) sample before and after K dosing. 
	(c), (d) Fermi surface maps for the as-grown PLCCO ($x=0.08$) samples \#1 and \#2, respectively. 
	The results of tight-binding fitting (eq.~\ref{TB_AFPG}) are shown as blue curves. 
	EDCs (e) at the node and (f) at the hot spot taken from the points indicated in (c), (d), and Fig.~\ref{character}(b). 
	} 
	\label{Kdosed}
\end{figure*}
K deposition was carried out on the surface of two distinct as-grown PLCCO ($x=0.08$) samples 
\#1 and \#2. Upon dosing, alkali-metal atoms adsorb on the surface and interact with the 
substrate material. In general, the interaction is ionic at low coverage, where alkali metal is 
partially ionized and electrons are injected into the surface layers of the sample~\cite{GurneyPhysRev1935,DiehlJPCM1997,stampfl1995theory} [Fig.~\ref{Kdosed}(a)]. 
The inelastic mean free path of photoelectrons  traveling inside PLCCO with the kinetic energy 
of 50~eV is estimated to be $\sim 5$~\AA~\cite{TPP}, while the CuO$_2$-layer spacing of PLCCO is 
6.1~\AA~\cite{guguchia2017pressure}. ARPES measurements conducted at $h\nu$ = 55~eV thus predominantly detects the electronic 
structure of the CuO$_2$ plane closest to the surface, where electron concentration is tuned 
by the amount of adsorbed alkali-metal atoms. At the initial stage of deposition, electron 
doping level increases accordingly with the increase of coverage. Concomitantly, however, the 
distance between the adsorbed alkali-metal atoms decreases, and in turn electrostatic repulsion increases 
due to the positive charge of the adsorbates. In order to reduce the total energy by weakening 
the repulsion, a fraction of the valence electrons begins to flow back from the sample to the alkali 
metal at a certain threshold coverage level~\cite{GurneyPhysRev1935,DiehlJPCM1997,stampfl1995theory}. The amount 
of electrons that can be doped to the sample is thus limited, and one needs to make several 
attempts of alkali-metal dosing and ARPES characterizations to identify the limit and to adjust 
the condition. In Fig.~\ref{Kdosed}(b), core-level spectra of sample~\#1 before and after K 
dosing to achieve the maximum electron doping are presented. A K 3$p$ peak develops once K is 
dosed, assuring the adsorption of K on the sample surface. No appreciable splitting in the 
K~3$p$ peak implies the coverage less than one monolayer~\cite{KimScience2014,KyungnpjQM2021}. 
After alkali-metal deposition, the charge balance between the sample and alkali metal could be 
sensitively modified by the adsorption of residual gas molecules~\cite{WeimerSS1985}. ARPES 
measurements of the K-deposited surface were therefore conducted within 1.5 hours after dosing, 
while the stability of the electron-doping level is secured.\\
\indent Considering the limited lifetime of the electron-doped surface state, we focused on acquiring a 
Fermi surface map and a high statistics energy-momentum map along a single cut, either through 
the node or the hot spot, for each sample. As shown in Figs.~\ref{Kdosed}(c) and (d), an 
unreconstructed, round-shaped Fermi surface centered at the BZ corner recovers for both samples. 
We fitted the Fermi surface to the two-dimensional tight binding model (eq.~\ref{TB_AFPG}) with $\Delta = 0$~eV, 
and from their area the electron concentrations of $n_\mathrm{FS} = 0.249 \pm 0.005$ and $0.209 \pm 0.006$ per Cu atom 
were estimated for samples \#1 and \#2, respectively. The increase of the electron concentrations 
with respect to the pristine surface [Fig.~\ref{character}(b)] should result from electron doping by adsorbed K atoms. 
The concentrations of additionally doped electrons are comparable to the highest values reported 
for alkali-metal deposited transition-metal oxides such as cuprates~\cite{HossainNature2008,HuNatCommun2021} 
and iridates~\cite{KimScience2014}. The difference in $n_\mathrm{FS}$ between the present 
two samples is likely due to instabilities in the alkali-metal flux arising from the fluctuating 
condition of dispensers. At the node, the gap observed on the pristine surface is 
fully suppressed [Fig.~\ref{Kdosed}(e)]. Spectral intensity near the Fermi level is enhanced also at the 
hot spot [Fig.~\ref{Kdosed}(f)], but spectral-weight depletion at the Fermi level is still significant as one can 
see from the leading-edge midpoint of the EDC positioned below the Fermi level.\\
\indent Acquiring reference ARPES data for the reduction-annealed sample would allow for a solid 
comparison on the effects of electron doping and oxygen reduction. To that end, we measured the 
annealed, superconducting PLCCO ($x=0.10$) sample. A Fermi surface without any reconstruction 
was observed~[Fig.~\ref{annealed}(a)], which can be fitted to the tight-binding model with 
$\Delta=0$~eV like the case of the K-dosed as-grown samples. The electron concentration 
$n_\mathrm{FS}$ estimated from Fermi surface area amounts to $0.145 \pm 0.004$. This is significantly larger 
than the Ce concentration but smaller than that of the K-dosed as-grown PLCCO ($x=0.08$) samples. 
The difference in $n_\mathrm{FS}$ between the annealed and K-dosed as-grown samples can be directly 
seen in the plot of Fermi wavenumber $k_\mathrm{F}$ [Fig.~\ref{annealed}(b)]; the area of the 
hole-like Fermi surface centered at the BZ corner is larger for the annealed sample. The contrast 
is also evident in the momentum distribution curve at the Fermi level [Fig.~\ref{annealed}(c)] 
plotted along the BZ boundary [black arrow in Fig.~\ref{annealed}(a)]. The separation between the 
two antinodal segments is smaller for the annealed sample, reflecting the larger size of the 
hole-like Fermi surface. Despite the clear evidence that the annealed sample has a smaller 
electron concentration, the spectral-weight recovery at the hot spot is even more pronounced as 
shown in Fig.~\ref{annealed}(d). In fact, any discernible signature of gap opening disappears 
after reduction annealing. \\

\section{Discussion}
\begin{figure}[ht]
	\begin{center}
		\includegraphics[width=0.48\textwidth]{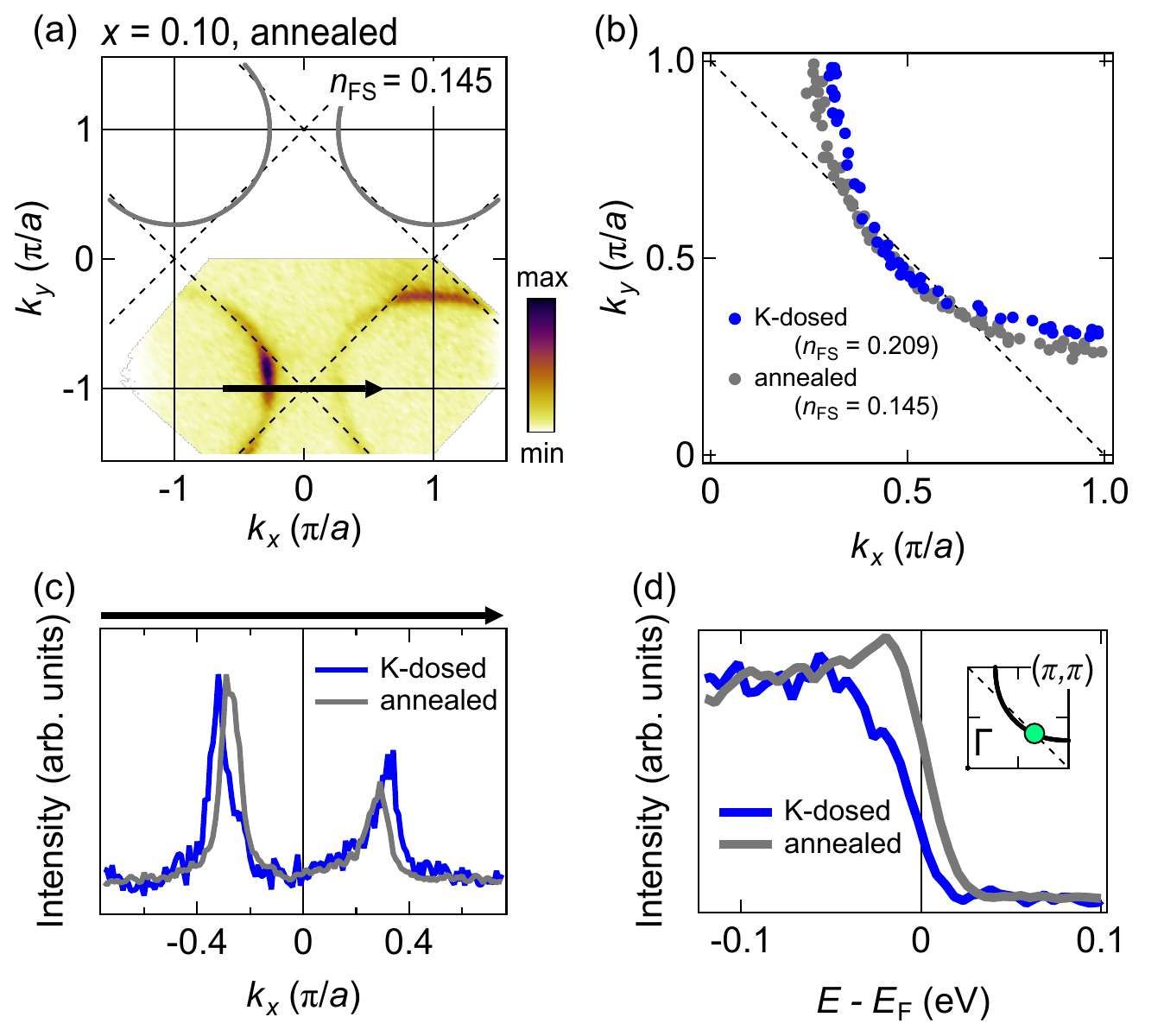}
	\end{center}
	\caption{\textbf{Influence of disorder on the pseudogap.} (a) Fermi surface map of the annealed PLCCO 
	($x=0.10$) sample. The black curve indicates the result of tight-binding fitting (eq.~\ref{TB_AFPG}). 
	(b) Comparison of Fermi wavenumber $k_\mathrm{F}$ for the $x=0.10$ annealed sample and the K-dosed 
	$x=0.08$ as-grown sample. (c) Momentum distribution curves for the two samples at 
	the Fermi level plotted along the BZ boundary [black arrow in (a)]. To alleviate intensity variation due 
	to matrix-element effects, photoemission signals have been symmetrized with respect to the $k_x+k_y=0$ line. 
	(d) EDCs at the hot spot (indicated in the inset) for the two samples. 
	} 
	\label{annealed}
\end{figure}
To clarify how electron doping and oxygen reduction relate to the emergence of superconductivity, 
the most straightforward approach is to directly detect a superconducting gap. However, the 
superconducting gap of electron-doped cuprates is as small as $\sim 5$~meV~\cite{MatsuiPRL2005,XuNatPhys2023,HorioPRB2019d-wave}. 
For the completion of measurements within the limited lifetime of the K-dosed surface state, 
photoelectron count rate needs to be prioritized at the expense of energy resolution, making 
the direct observation of the small superconducting gap quite challenging. Instead, in the 
current work, we focus on the pseudogap whose energy scale is one order of magnitude larger than 
that of the superconducting gap. As the long-range AF order of electron-doped cuprates is 
suppressed, a spin-glass-like, short-range AF state appears~\cite{KuroshimaPhysicaC2003}. 
The short-range AF correlation is known to persist even in the superconducting regime
~\cite{MotoyamaNature2007}. By ARPES studies on such samples with short-range AF correlation, 
a spectral-weight suppression -- so-called pseudogap -- was found around the hot spot
~\cite{ArmitagePRL2001,MatsuiPRB2007,ParkPRB2007,RichardPRL2007,IkedaPRB2009,SongPRB2012,ParkPRB2013,SongPRL2017,XuNatPhys2023,HorioPNAS2025,SongNatCommun2025}. 
While short-range AF order or fluctuations have been considered to be responsible for the 
pseudogap formation~\cite{ParkPRB2007,ParkPRB2013,SongPRL2017,XuNatPhys2023,SongNatCommun2025}, 
a recent study~\cite{HorioPNAS2025} put forward strong electron correlation that produces 
short-range AF spin and charge correlations as the fundamental source of the pseudogap. While 
the exact origin is yet to be fully clarified, the competition between the pseudogap and superconductivity 
has been established~\cite{RichardPRL2007,HorioNatCommun2016,SongPRL2017}. Hence, one can 
indirectly derive information linked to superconductivity by monitoring the changes in the 
pseudogap.\\
\indent The transition of the Fermi surface shape by K dosing, from the reconstructed electron-like one 
to the unreconstructed hole-like one~[Figs.~\ref{Kdosed}(c) and (d)], suggests the suppression of long-range 
AF order triggered by electron doping from $n_\mathrm{FS} = 0.079$ to $0.209$. 
However, spectral weight depletion near the Fermi level remains significant around 
the hot spot [Fig.~\ref{Kdosed}(f)], indicating the persistence of the pseudogap. This implies 
that, even after considerable amount of electron doping, short-range AF spin and charge correlations 
survive and inhibit superconductivity. In contrast, the pseudogap virtually disappears after 
efficient reduction annealing [Fig.~\ref{annealed}(d)]. It should be noted that the electron concentration of 
the present reduction annealed sample ($n_\mathrm{FS} = 0.145$) is substantially smaller than that of the 
K-dosed as-grown sample ($n_\mathrm{FS} = 0.209$). Therefore, the pseudogap -- and hence superconductivity -- 
of electron-doped cuprates cannot be described solely by electron concentration; the pseudogap 
is reinforced by the presence of apical oxygen impurities.

With the increase of itinerant electrons, electron correlation is screened and effectively 
weakened. AF correlation is also disturbed through the dilution of spins on the Cu sites. On the 
other hand, as pointed out in previous studies~\cite{XuPRB1996}, apical oxygen impurities act as 
a strong scattering center that causes localization of itinerant carriers. Electron localization 
around the impurity should lead to the recovery of short-range charge and AF spin correlations 
around the impurity site~\cite{AdachiJPSJ2016,AdachiCondMat2017}, strengthening the pseudogap. Although the addition 
of electrons could enhance the screening of the disorder potential and alleviate the pseudogap opening, 
the present results suggest that it is not so effective as removing the impurity apical oxygen atoms 
by reduction annealing. Note that reduction annealing should also create oxygen defects in the CuO$_2$ planes and/or the 
(Pr,La,Ce)$_2$O$_2$ layers, yielding an electron concentration larger than the Ce concentration. Considering the 
strong suppression of the pseudogap after reduction annealing, however, the disorder potential 
of those defects is substantially smaller than that of the apical-site impurities. This implies 
that the created defects mostly reside in the (Pr,La,Ce)$_2$O$_2$ layer, which is located farther 
from the CuO$_2$ plane than the apical site and hence less influential to the electrons in the 
CuO$_2$ plane [see Fig.~\ref{character}(a)]~\cite{EisakiPRB2004}. Consequently, reduction annealing effectively suppresses 
electron localization as well as the psuseudogap, allowing for the emergence of superconductivity. 
The present work thus demonstrates the crucial role of oxygen reduction that cannot be replaced 
by electron doping. For comprehensive understanding on the intrinsic physics of electron-doped 
cuprates, such differentiation of the disorder contributions is of key importance.

\section{Conclusions}
To summarize, we deposited K on the as-grown electron-doped cuprate PLCCO ($x=0.08$) and 
conducted ARPES measurements to study the influence of electron doping without modifying the 
oxygen content in the sample. While the AF reconstruction of the Fermi surface was suppressed 
after K deposition, spectral-weight depletion at the hot spot due to pseudogap opening persisted 
even at the electron doping level of $n_\mathrm{FS} = 0.209$. On the other hand, efficiently 
reduction-annealed samples with the electron concentration smaller than that of the K-dosed 
as-grown samples did not show a discernible signature of pseudogap. 
The results highlight the persistence of a strong disorder potential in as-grown electron-doped 
cuprates that cannot be screened out by electron doping and can only be eliminated through 
efficient oxygen reduction.

\section{Acknowledgements}
This work was supported by JSPS KAKENHI Grant Numbers JP19H01841, JP21K13872, JP24KK0061 and JP25K00218. 
The ARPES measurements were performed at the BL5U of UVSOR Synchrotron Facility, Institute for 
Molecular Science (IMS program 22IMS6839, 23IMS6652, and 24IMS6644).

%

\end{document}